\documentclass[pra,twocolumn,aps,floatfix,showpacs,tightenlines,superscriptaddress,amsmath,amssymb]{revtex4}
\usepackage{amssymb,amsbsy,epsfig,color,graphicx}

\begin{document}

\title{Realistic continuous-variable quantum teleportation with non-Gaussian resources}

\author{F. Dell'Anno}
\affiliation{Dipartimento di Matematica e Informatica,
Universit\`a degli Studi di Salerno, Via Ponte don Melillo,
I-84084 Fisciano (SA), Italy} \affiliation{CNR-INFM Coherentia, Napoli,
Italy, CNISM, and INFN Sezione di Napoli, Gruppo collegato di Salerno,
Baronissi (SA), Italy}

\author{S. De Siena}
\affiliation{Dipartimento di Matematica e Informatica,
Universit\`a degli Studi di Salerno, Via Ponte don Melillo,
I-84084 Fisciano (SA), Italy} \affiliation{CNR-INFM Coherentia,
Napoli, Italy, CNISM, and INFN Sezione di Napoli, Gruppo collegato di
Salerno, Baronissi (SA), Italy}

\author{F. Illuminati}
\affiliation{Dipartimento di Matematica e Informatica,
Universit\`a degli Studi di Salerno, Via Ponte don Melillo,
I-84084 Fisciano (SA), Italy} \affiliation{CNR-INFM Coherentia,
Napoli, Italy, CNISM, and INFN Sezione di Napoli, Gruppo collegato di
Salerno, Baronissi (SA), Italy} \affiliation{ISI Foundation for
Scientific Interchange, Viale Settimio Severo 65, 00173 Torino,
Italy} \affiliation{Corresponding author. Electronic address:
illuminati@sa.infn.it}

\date{October 14, 2009}

\begin{abstract}
We present a comprehensive investigation of nonideal continuous-variable
quantum teleportation implemented with entangled non-Gaussian resources.
We discuss in a unified framework the main decoherence mechanisms, including
imperfect Bell measurements and propagation of optical fields in lossy fibers,
applying the formalism of the characteristic function. By exploiting
appropriate displacement strategies, we compute analytically the success
probability of teleportation for input coherent states, and two classes of
non-Gaussian entangled resources: Two-mode squeezed Bell-like states
(that include as particular cases photon-added and photon-subtracted
de-Gaussified states), and two-mode squeezed cat-like states.
We discuss the optimization procedure on the free parameters of the
non-Gaussian resources at fixed values of the squeezing and of the
experimental quantities determining the inefficiencies of the non-ideal protocol.
It is found that non-Gaussian resources enhance significantly the
efficiency of teleportation and are more robust against decoherence than
the corresponding Gaussian ones. Partial information on the alphabet of
input states allows further significant improvement in the performance
of the nonideal teleportation protocol.
\end{abstract}


\maketitle

\section{Introduction}

In recent years, theoretical and experimental investigation has been focused
on the engineering of highly nonclassical, non-Gaussian states of the radiation
field (for a review, see e.g. \cite{PhysRep}).
Interest in the production of non-Gaussian optical states is due to their
strongly nonclassical properties, such as entanglement and negativity
of the quasi-probability phase-space distributions,
that are important for the efficient implementation of quantum information and
communication protocols \cite{PhysRep,KimBS,KitagawaPhotsub,DodonovDisplnumb,Cerf}, and for quantum estimation tasks \cite{QEstimNoi}.

Several schemes for the generation of non-Gaussian states, both single-mode and two-mode,
have already been proposed \cite{CxKerrKorolkova,AgarTara,DeGauss1,DeGauss2,DeGauss3,DeGauss4,DeGauss5},
and many successful and encouraging experimental realizations have been reported recently
\cite{ZavattaScience,ExpdeGauss1,ExpdeGauss2,Grangier,BelliniProbing,GrangierCats}.
In line of principle, a very important result concerns the rigorous proof that various
nonclassical properties are minimized by Gaussian states \cite{ExtremalGaussian}.
Therefore, it is reasonable to expect that the use of non-Gaussian resources
may improve significantly the performance of quantum information protocols.
In particular, concerning quantum teleportation with continuous variables (CV),
it has been shown that the success probability of teleportation can be greatly increased
by using entangled non-Gaussian resources in the framework of the ideal Braunstein-Kimble (B-K)
protocol \cite{KitagawaPhotsub,Opatrny,Cochrane,Olivares,CVTelepNoi,YangLi}.

Indeed, it has been shown that some specific two-mode non-Gaussian states,
dubbed squeezed Bell-like states (that include as subcases photon-added and
photon-subtracted de-Gaussified states \cite{CVTelepNoi}), when used as entangled
resources provide a significant increase in the teleportation fidelity
of single-mode input states under the ideal protocol \cite{CVTelepNoi}.
Such an enhancement is due to a balancing of three different features \cite{CVTelepNoi}:
The entanglement content of the resources, their (appropriately defined)
degree of affinity with the two-mode squeezed vacuum, and their (suitably measured)
amount of non-Gaussianity. For the precise definition of the last two quantities, see Refs.~\cite{CVTelepNoi,GenoniNonGaussy}). It has been suggested \cite{CVTelepNoi}
that such states can be produced by combining simultaneous phase-matched multiphoton
processes and conditional measurements.

The analysis of Ref.~\cite{CVTelepNoi} has been later extended to consider other
classes of non-Gaussian resources, such as two-mode squeezed symmetric superpositions
of Fock states and of squeezed cat-like states, that allow high levels of performance
in the teleportation of single-mode input states \cite{CVTelepNoisyNoi}.
A partial preliminary analysis of non-ideal cases has also been performed
\cite{CVTelepNoisyNoi} by considering simple superpositions of independently
generated fields converging on a common spatial volume, such as superpositions
of a two-mode pure non-Gaussian resource and a two-mode thermal state \cite{Glauber}.
In this elementary instance, mixed non-Gaussian entangled states remain preferred
resources for teleportation when compared to mixed twin-beam Gaussian states \cite{CVTelepNoisyNoi}.
\\
In this work, using the formalism of the characteristic function, we study in full generality
the Braunstein-Kimble protocol for CV teleportation in realistic conditions and with
non-Gaussian entangled resources. We include in our investigation the main sources of
decoherence that lead to the degradation of the transferred quantum information,
such as losses due to imperfect homodyne measurements, and damping due to the
propagation of the optical fields in lossy fibers. The effects of these inefficiencies
have already been considered, among others, in Refs.~\cite{VukicsnonidealTelep,TelepChizhov}.
In particular Ref.~\cite{VukicsnonidealTelep} is concerned with the study of imperfect
Bell measurements, while in Ref.~\cite{TelepChizhov} the authors investigate
the limits of quantum teleportation due to photon absorption during propagation in fibers.
Besides considering each problem separately, these and related works are always restricted
to the use of Gaussian resources. The main object of the present work is to investigate the
effect of the simultaneous presence of all sources of imperfection on the performance of
CV teleportation protocols with non-Gaussian resources, and their robustness against decoherence.

A general and exhaustive analysis turns out to be possible in the framework
of the characteristic function representation. This method has been
discussed in full generality for the description of ideal CV teleportation
\cite{MarianCVTelep}, and applied first to the case of Gaussian \cite{MarianCVTelep}
and non-Gaussian resources \cite{CVTelepNoi,CVTelepNoisyNoi}.
We will then extend the formalism to include the description of nonideal CV teleportation,
including realistic Bell measurements and decoherence due to propagation in noisy channels.
In order to investigate different optimization strategies of the nonideal protocol,
we will discuss optimization over the free parameters of the non-Gaussian resources
as well as over the gain factor associated with the transmitted
classical information \cite{TelepGainBowen} (for various strategies of gain tuning and
of optimal gain see also Refs.~\cite{TelepChizhov,TelepIde,TelepTailored}).
Indeed, in the instance of non-Gaussian resources, the gain can be considered
as a further free parameter suitable for optimization.

The paper is organized as follows.
In section~\ref{SecCharFuncTelep} we extend the characteristic function formalism
to include the case of realistic CV quantum teleportation.
In section~\ref{SecEntangRes} we introduce and discuss the main properties
of some classes of non-Gaussian entangled resources.
In section~\ref{SecUnityGain} we study the efficiency of the quantum teleportation protocol
in the instance of fixed given values of the gain.
In section~\ref{SecNonunityGain} we carry out an optimization procedure of the protocol
over the entangled resources and the gain parameter.
Finally, in section \ref{secConclusions} we draw our conclusions and discuss some
outlook on current and future research.

\section{Nonideal CV teleportation protocol in the characteristic function formalism}
\label{SecCharFuncTelep}
\begin{figure}[t]
\centering
\includegraphics*[width=8.5cm]{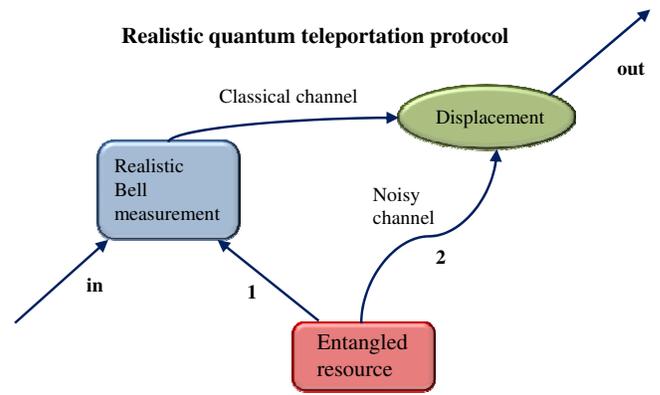}
\caption{(Color online) Pictorial representation of the nonideal B-K CV quantum teleportation
protocol: In the first step, the input mode is mixed by Alice with one
of the two beams (modes) of the entangled resource; the ensuing state
is then subject to a realistic Bell measurement. The result of the measure
is communicated to Bob through a classical channel. In the second step,
a unitary transformation, determined by the previous measurement, is applied to the second
mode of the entangled resource, that is affected by decoherence during the propagation
in a noisy channel, e.g. a lossy fiber. The ensuing output state is the final
teleported state.}
\label{FigRealQuantTel}
\end{figure}

In this section, we describe the realistic B-K CV teleportation protocol
in the formalism of the characteristic function. Although several alternative formalisms
are available for the description of the B-K CV teleportation protocol
\cite{BraunsteinKimble,TelepFormal1,TelepFormal2,TelepFormal3,FuruRep,vanLoockTelep},
the characteristic function representation proves to be particularly convenient
when considering the nonideal case and non-Gaussian resources.
The description of nonideal teleportation requires the introduction
of mechanisms of loss and inefficiency in the main steps of the protocol.
Indeed, a schematic description of the nonideal protocol is depicted in Fig.~\ref{FigRealQuantTel}.

\begin{figure}[t]
\centering
\includegraphics*[width=8.5cm]{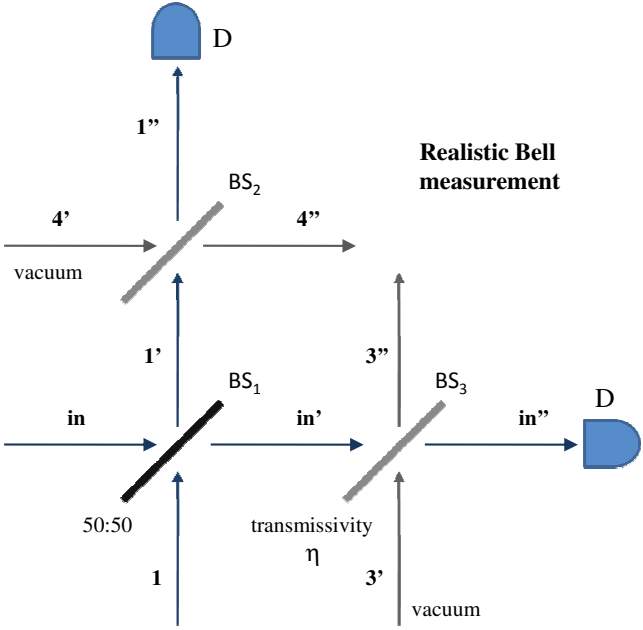}
\caption{(Color online) Model scheme of a realistic Bell measurement.
The model takes into account the non-unity efficiency of
the detectors $D$ performing the homodyne measurement.
In the depicted scheme, such inefficiency is simulated by the
introduction of two fictitious beam splitters $BS_{2}$ and $BS_{3}$,
with equal transmissivity $\eta$.}
\label{FigRealQuantTel2}
\end{figure}

The input state and the entangled resource states are assumed to be initially pure.
This is not a serious limitation because one can always map the case of
a nonideal teleportation protocol with noisy (mixed) inputs and resources
to an equivalent protocol with pure inputs and resources but with a
correspondingly larger amount of noise affecting the protocol. A simple
illustrative example of this equivalence will be discussed later on
in the present work.

The single-mode input state ($in$) is mixed to mode $1$
of the entangled resource at a beam splitter. At the first user's
(Alice) location, a Bell measurement, consisting in homodyne detections,
is performed on the obtained state of mode $in$ and mode $1$.
In order to describe a nonideal measurement, one needs
to model/simulate the inefficiencies of the photo-detectors.
A realistic detector can be modeled by placing a fictitious beam splitter,
i.e. a partly transmitting mirror, in front of an ideal detector \cite{LeonhardtRealHomoMeasur}.
Based on such a prescription, a scheme describing a realistic Bell measurement
is shown in Fig.~\ref{FigRealQuantTel2}.

After a realistic Bell measurement,
the result is transmitted to the receiver (Bob) through a classical channel.
The mode $2$ of the entangled resource propagates in a noisy channel, as a lossy
fiber, to Bob's location, where it undergoes a unitary displacement
according the result of the Bell measurement It is important to note that the
generation of the entangled resource can take place close to the sender, and
typically very far away from the receiver, as one of the main tasks of quantum
teleportation is the transfer of quantum information across long distances.
It is then legitimate to assume that the radiation field associated
to mode $1$ is not affected by losses due to propagation, while the field
associated to mode $2$, which usually has to propagate over much longer distances,
can be strongly affected by decoherence. The degradation of quantum information
is caused by the propagation of field mode $2$ in a noisy channel.
Therefore, the output teleported state depends both on the
inefficiency of the homodyne detectors and the decoherence rate
of the noisy channel. \\

We can formalize the effects of the above-described dynamics as follows.
Let us denote by
$\rho_{in} \,=\, |\phi\rangle_{in}\,_{in}\langle \phi|$
and $\rho_{res} \,=\, |\psi\rangle_{12}\,_{12}\langle \psi|$
the density matrices associated, respectively, with the
single-mode pure input state and with the two-mode pure
entangled resource. The single-mode input state is initially
disentangled from the two-mode entangled resource, so that the
the initial three-mode field is
$\rho_{0} \,=\, \rho_{in} \otimes \rho_{res}$,
and the initial global characteristic function reads
\begin{eqnarray}
&&\chi_{0}(\alpha_{in};\alpha_{1};\alpha_{2}) \,=\,
Tr[\rho_{0} \; D_{in}(\alpha_{in})\,
D_{1}(\alpha_{1})\, D_{2}(\alpha_{2})] \nonumber \\
&& \nonumber \\
&&\,=\, \chi_{in}(\alpha_{in})\; \chi_{res}(\alpha_{1};\alpha_{2})  \,,
\label{globcharfuncinitial}
\end{eqnarray}
where $Tr$ denotes the trace operation, $D_{j}(\alpha_{j})$ denotes the
displacement operator for the mode $j$ ($j=in,1,2$), $\chi_{in}$ is the
characteristic function of the input state, and $\chi_{res}$ is the
characteristic function of the entangled resource.
By defining the quadrature operators $X_{j} \,=\,
\frac{1}{\sqrt{2}}(a_{j}+ a_{j}^{\dag})$ and $P_{j} \,=\,
\frac{i}{\sqrt{2}}(a_{j}^{\dag}- a_{j})$ $(j=in,1,2)$, and the associated
phase-space real variables $x_{j} \,=\,
\frac{1}{\sqrt{2}}(\alpha_{j}+ \alpha_{j}^{*})$ and $p_{j} \,=\,
\frac{i}{\sqrt{2}}(\alpha_{j}^{*}- \alpha_{j})$,
the characteristic function can be written in terms of $x_{j}$, $p_{j}$, i.e.
$\chi_{0}(\alpha_{in};\alpha_{1};\alpha_{2}) \equiv \chi_{0}(x_{in},p_{in};x_{1},p_{1};x_{2},p_{2})$.
\\
The first step of the protocol consists in the Bell measurement at Alice's location,
that is the homodyne measurements of the first quadrature of the mode $1$
and of the second quadrature of the mode $in$, with results $\tilde{x}$ and $\tilde{p}$, respectively.
After such nonideal Bell measurement, the remaining mode $2$ is left in a mixed state described by
the corresponding single-mode characteristic function $\chi_{Bm}(x_{2},p_{2})$. In appendix \ref{AppendixRealTel} we prove in detail how to compute it in full generality for arbitrary single-mode inputs and arbitrary two-mode entangled resources. Here we report only the final expression:
\begin{eqnarray}
&&\chi_{Bm}(x_{2},p_{2}) = \frac{\mathcal{P}^{-1}(\tilde{p},\tilde{x})}{(2\pi)^{2}}
\int d\xi d\upsilon \;  e^{i \xi\tilde{p}-i \tilde{x} \upsilon} \times \nonumber \\
&& \nonumber \\
&& \chi_{in}\left(\frac{T \xi}{\sqrt{2}}\,,\frac{T \upsilon}{\sqrt{2}} \right)
\chi_{res}\left(\frac{T \xi}{\sqrt{2}}\,,-\frac{T \upsilon}{\sqrt{2}};x_{2},p_{2}\right)\times
\nonumber \\
&& \nonumber \\
&&\exp\left\{-\frac{R^{2}}{4}( \xi^{2}+ \upsilon^{2}) \right\} \,,
\label{chiBellMeasurement}
\end{eqnarray}
where $T^2$ and $R^2=(1-T^{2})$ denote, respectively, the transmissivity and reflectivity
of the beam splitters that model the losses.
The function $\mathcal{P}(\tilde{p},\tilde{x})$ is the distribution of the measurement outcomes
$\tilde{p}$ and $\tilde{x}$ (see appendix \ref{AppendixRealTel}).
Note that the Gaussian exponential in Eq.~(\ref{chiBellMeasurement})
is related to the vacua entering the input ports of the fictitious beam splitters.
\\
Afterwards, mode $2$ propagates in a damping channel, like, e.g., a lossy fiber,
before it reaches Bob's location. The Markovian dynamics of a system subject to
damping is described, in the interaction picture, by the following master equation
for the density operator $\rho$ \cite{WallsMilburn,DecohReview}:
\begin{equation}
\partial_{t} \rho \,=\, \frac{\Upsilon}{2}
\left\{ n_{th} L[a_{2}^{\dag}] \rho + (n_{th}+1) L[a_{2}] \rho \right\} \,,
\label{MasterEq}
\end{equation}
where the Lindblad superoperators are defined as $L[\mathcal{O}]
\rho \equiv 2 \mathcal{O} \rho \mathcal{O^{\dag}}
- \mathcal{O^{\dag}} \mathcal{O} \rho - \rho \mathcal{O^{\dag}} \mathcal{O}$,
$\Upsilon$ is the mode damping rate, and $n_{th}$ is the number of thermal photons.
Finally, at Bob's location, a displacement $\lambda= g(\tilde{x}+i \tilde{p})$
is performed on mode $2$. The real parameter $g$ is the so-called gain factor
\cite{vanLoockTelep}.
The combined effect of propagation in a damping channel and unitary displacement
determines the characteristic function $\chi_{out}(x_{2},p_{2})$ of the final
output state of the teleportation protocol (see appendix \ref{AppendixRealTel}
for details):
\begin{eqnarray}
&&\chi_{out}(x_{2},p_{2}) \,=\,
\chi_{in}\left(g T x_{2}\,,g T p_{2} \right) \times \nonumber \\
&& \nonumber \\
&& \chi_{res}\left(g T x_{2}\,,-g T p_{2};e^{-\frac{\tau}{2}}x_{2},e^{-\frac{\tau}{2}}p_{2}\right) \times \nonumber \\
&& \nonumber \\
&&\exp\left\{-\frac{1}{2} \Gamma_{\tau,R}(x_{2}^{2}+p_{2}^{2})\right\} ,
 \label{chioutfinale}
\end{eqnarray}
where $\tau=\Upsilon t$, and the thermal "renormalized" phase-space
covariance $\Gamma_{\tau,R}$ is defined as:
\begin{equation}
\Gamma_{\tau,R} \,=\, (1-e^{-\tau})\left(\frac{1}{2}+n_{th}\right)+g^{2}R^{2} \,.
\label{Gammadef}
\end{equation}
The form of Eq.~(\ref{chioutfinale}) highlights the different roles played
by the two sources of noise introduced in the teleportation protocol, associated,
respectively, to the damping rate $\Upsilon$ and the reflectivity $R^2$.
The two decoherence mechanisms act separately but also in combination, as
one can see from the Gaussian exponential factor in Eq.~(\ref{chioutfinale}),
which in fact is nonvanishing for $R \neq 0$ and/or $\tau \neq 0$.
The effect of the imperfect Bell measurement is expressed also
by the presence of the scale factor $T$ in the arguments of the input and
resource characteristic functions $\chi_{in}$ and $\chi_{res}$. Viceversa,
decoherence due to noisy propagation affects obviously only mode $2$
by means of the exponentially decreasing weight $e^{-\frac{\tau}{2}}$
in the arguments of $\chi_{res}$. The factorized form of the output
characteristic function, holding for the ideal protocol \cite{MarianCVTelep},
\begin{equation}
\chi_{out}(x_{2},p_{2})=\chi_{in}(x_{2},p_{2})\,\chi_{res}(x_{2},-p_{2};x_{2},p_{2}),
\label{MarianFormula}
\end{equation}
is recovered, as expected, from Eq.~(\ref{chioutfinale}) when $R=0$ $(T=1)$, $\Upsilon=0$ $(\tau=0)$,
and $g=1$.

\section{Entangled resources: Two classes of optimized non-Gaussian states}
\label{SecEntangRes}
Given the general description of the nonideal protocol in terms of the characteristic functions,
in this section we analyze the performance of two classes of non-Gaussian entangled
resources for the teleportation of input coherent states, respectively, the two-mode
squeezed Bell-like states $|\psi\rangle_{SB}$ and the two-mode squeezed cat-like
states $|\psi\rangle_{SC}$:
\begin{equation}
|\psi\rangle_{SB} = S_{12}(\zeta)
\{\cos\delta |0,0 \rangle + e^{i \theta} \sin\delta
|1,1 \rangle \} ,
\label{squeezBell}
\end{equation}
\begin{eqnarray}
&&|\psi\rangle_{SC} = \mathcal{N}_{SC} S_{12}(\zeta)
\{\cos\delta |0,0 \rangle + e^{i \theta} \sin\delta
|\gamma,\gamma \rangle \} , \nonumber \\
\label{squeezCat}
\end{eqnarray}
where $S_{12}(\zeta) = e^{ -\zeta a_{1}^{\dag}a_{2}^{\dag} + \zeta
a_{1}a_{2}}$ is the two-mode squeezing operator, $\zeta=r e^{i\phi}$,
$|m \, , n \rangle  \equiv |m \rangle_{1} \otimes |n \rangle_{2}$
is a two-mode Fock state (of modes 1 and 2),
$|\gamma,\gamma \rangle\equiv |\gamma\rangle_{1}\otimes |\gamma\rangle_{2}$
is a symmetric two-mode coherent state with complex amplitude $\gamma =|\gamma| e^{i \varphi}$,
and the normalization factor $\mathcal{N}_{SC}$ is
$\mathcal{N}_{SC}=\{1+ e^{-|\gamma|^{2}}\sin 2\delta \cos\theta\}^{-1/2}$.
In order to obtain a maximization of the teleportation fidelity, it is necessary
to perform a simultaneous balanced optimization, on the free parameters
($\delta$, $\theta$, $\gamma$), of some partially competing properties
\cite{CVTelepNoi,CVTelepNoisyNoi}. These include the entanglement content,
the amount of non-Gaussianity of the state \cite{GenoniNonGaussy}, and a
squeezed-vacuum-affinity $\mathcal{G}$. For a generic pure state $|\psi\rangle$,
the latter is defined as \cite{CVTelepNoi,CVTelepNoisyNoi}:
\begin{equation}
\mathcal{G} = \sup_{r} |\langle -r|\psi\rangle|^{2} \, ,
\end{equation}
with $|-r\rangle = S_{12}(-r)|0,0\rangle$.
Indeed, the optimal non-Gaussian resources (\ref{squeezBell}) and (\ref{squeezCat})
exhibit a sufficient squeezed-vacuum-affinity, which then appears to be a crucially
needed property in order to select efficient and highly performing non-Gaussian resources.
For instance, one could as well consider a different form of squeezed Bell-like state,
the so-called "Buridan donkey" or "Hamlet" state $|\psi\rangle_{SB2}$, that is
obtained from the singlet Bell state as follows:
\begin{equation}
|\psi\rangle_{SB2} = S_{12}(\zeta)\{\cos\delta |0,1 \rangle + e^{i \theta} \sin\delta|1,0 \rangle \} .
\label{squeezBell2}
\end{equation}
It is indeed simple to verify that, although such a state, at fixed squeezing,
is more entangled than a Gaussian twin beam, it performs less efficiently both
in the ideal and in the realistic teleportation protocol.
This fact can be understood if one looks at the behavior of the squeezed vacuum affinity \cite{CVTelepNoi}.
Namely, the Buridan donkey state Eq.~(\ref{squeezBell2}) does not contain the fundamental Gaussian 
contribution coming from the squeezed vacuum.
Therefore, it is ``unbalanced``, in the sense that it is less affine to the squeezed vacuum,
and excessively non-Gaussian compared to the optimized Bell-like state $|\psi\rangle_{SB}$.
Therefore, the fine interplay among these three quantities,
i.e. the entanglement, the degree of non-Gaussianity, and, in particular, the squeezed vacuum affinity,
cannot be realized in the non-Gaussian resource (\ref{squeezBell2}) \cite{CVTelepNoi}.
The crucial role played by the squeezed vacuum affinity for the performance of
different non-Gaussian resources has been studied in Ref.~\cite{CVTelepNoisyNoi}.
In particular, it has been shown that, in the ideal teleportation protocol,
the two-mode squeezed symmetric superposition of Fock states,
i.e. $S_{12}(\zeta)\sum_{k=0}^{2}c_{k}|k,k \rangle$,
when optimized for the teleportation of both input coherent states and single-photon states,
reduces to a squeezed truncated twin beam,
i.e. $S_{12}(-r)\sum_{k=0}^{2}\tanh^{k} s|k,k \rangle$.
In the same paper, it has been also shown that, in the ideal teleportation protocol,
the optimized two-mode squeezed cat-like states, i.e. Eq.~(\ref{squeezCat}),
possess a high amount of squeezed vacuum affinity.
Therefore, besides a certain amount of entanglement, also a sufficient degree of
squeezed vacuum affinity appears to be necessary for a non-Gaussian resource to 
be optimal for a B-K teleportation protocol. 

Clearly, the performance of B-K teleportation protocols depends strongly 
on the structure of the second-order correlations in the entangled resources. 
In this sense the B-K protocol, with its structure of homodyne measurements, 
is particularly tailored to the use of Gaussian resources. 
Therefore, a non-Gaussian resource may improve on the performance 
of a corresponding Gaussian one only if the fundamental Gaussian 
contribution coming from the squeezed vacuum is subject to a not 
too drastic modification. A large value of the affinity assures 
that the non-Gaussian resource satisfies such a requirement. The
interplay of the affinity with the non-Gaussianity and the degree
of entanglement allows to single out those non-Gaussian resources
possessing higher-order correlations that add to the leading Gaussian structure 
of the two-mode entangled resource, thus enhancing further the protocol 
efficiency, and lacking those non-Gaussian contributions that are incompatible 
with the structure of the B-K protocol. 

Indeed, a very interesting question 
open for future investigation is the inverse of the one studied in the present
paper. Here we are analyzing the problem of optimizing non-Gaussian resources given the
B-K protocol. The inverse question would be that of adapting the protocol to the resources.
Namely, given a certain class of non-Gaussian squeezed resources with some 
given properties, one asks how the B-K protocol would
have to be modified in order to optimize the fidelity of teleportation.

We now proceed to determine the general expression of the fidelity of teleportation
in terms of the characteristic function for the three different non-Gaussian entangled
resources $|\psi\rangle_{SB}$, $|\psi\rangle_{SC}$, $|\psi\rangle_{SB2}$.
For instance, the two-mode characteristic function $\chi_{SB}$ of
the entangled resource (\ref{squeezBell}) reads
\begin{equation}
\chi_{SB}(\alpha_{1},\,\alpha_{2})=Tr[|\psi\rangle_{SB}\,_{SB}\langle\psi| \, D_{1}(\alpha_{1})D_{2}(\alpha_{2})] \, ,
\label{chiSB}
\end{equation}
and analogous expressions hold for $\chi_{SC}$ and $\chi_{SB2}$.
The corresponding explicit expression is obtained
using the two-mode Bogoliubov transformations
\begin{eqnarray}
&&S_{12}^{\dag}(\zeta)\, a_{i} \, S_{12}(\zeta)=\cosh r \, a_{i}
-e^{i\phi}\sinh r \, a_{j}^{\dag}, \, \nonumber \\
&& (i\neq j=1,2) \,,
\label{BogoliubovT}
\end{eqnarray}
and the relation
\begin{equation}
\langle m| D(\alpha) |n \rangle \,=\,
\left(\frac{n!}{m!}\right)^{1/2}\alpha^{m-n}e^{-\frac{1}{2}|\alpha|^{2}}
L_{n}^{(m-n)}(|\alpha|^{2}) \,,
\label{LaguerreFormula}
\end{equation}
where $L_{n}^{(m-n)}(\cdot)$ denotes the associated Laguerre polynomial of order $n$.

The quantity measuring the success probability of a teleportation
protocol is the fidelity of teleportation $\mathcal{F} \, = \,
Tr[\rho_{in}\rho_{out}]$. In the formalism of the characteristic function,
the fidelity reads
\begin{eqnarray}
\mathcal{F} =&&  \frac{1}{\pi} \int d^{2}\alpha \;
\chi_{in}(\alpha) \chi_{out}(-\alpha) \,,  \nonumber \\
&& \nonumber \\
&&\frac{1}{2\pi} \int dx_2 dp_2 \;
\chi_{in}(x_2, p_2) \chi_{out}(-x_2, -p_2) \,,
\label{Fidelitychi}
\end{eqnarray}
where $\alpha= \frac{1}{\sqrt{2}}(x_2 +i p_2 )$, $d^{2}\alpha = \frac{1}{2} dx_2 dp_2$,
and $\chi_{out}(\alpha)\equiv\chi_{out}(x_{2},p_{2})$ is given by Eq.~(\ref{chioutfinale}).
In the case of input coherent states $\rho_{in}=|\beta\rangle_{in}\,_{in}\langle\beta|$
with complex amplitude $\beta$, that we will always consider in the following,
the characteristic function of the input $\chi_{in}(\alpha)$ reads:
\begin{equation}
\chi_{in}(\alpha) \,=\,
e^{-\frac{1}{2}|\alpha|^{2}+(\alpha\beta^{*}-\alpha^{*}\beta)}
\; .
\label{chiCohin}
\end{equation}
Eq.~(\ref{Fidelitychi}) is the fundamental quantity that measures
the efficiency of a CV teleportation protocol (ideal or nonideal).
At fixed squeezing, the optimization procedure consists
in the maximization of the teleportation fidelity (\ref{Fidelitychi})
over the free parameters of the non-Gaussian entangled resources,
Eqs.~(\ref{squeezBell}), (\ref{squeezCat}), and (\ref{squeezBell2}).
Since it can be verified explicitly that the optimal choice for the
phases $\phi$, $\theta$, and $\varphi$ are
$\phi=\pi$ and $\theta=\varphi=0$, the squeezed Bell-like state $|\psi_{SB}\rangle$
and the Buridan donkey state $|\psi_{SB2}\rangle$ have a unique available
free parameter $\delta$. On the other hand, the squeezed cat-like state
$|\psi_{SC}\rangle$ has two free parameters, the angle $\delta$ and the
modulus $|\gamma|$. The analytical expressions of the teleportation fidelities
of input coherent states corresponding to the three different classes of non-Gaussian resources,
respectively $\mathcal{F}_{SB}^{(g)}(r,\delta)$, $\mathcal{F}_{SC}^{(g)}(r,\delta,|\gamma|)$,
and $\mathcal{F}_{SB2}^{(g)}(r,\delta)$ are reported in appendix~\ref{AppendixFid},
Eqs.~(\ref{FidelitySqBell}), (\ref{FidelitySqCat}), and (\ref{FidelitySqBell2}).
Let us notice that we have introduced the superscript $(g)$ to explicitly indicate
the dependence on the gain $g$. It is simple to verify that, for arbitrary $g$,
the fidelities are explicitly dependent on the amplitude $\beta$ of the input
coherent states. In the next sections, the (numerical) optimization procedures
of the fidelity $\mathcal{F}$ will be implemented following two different routes.
In section \ref{SecUnityGain} we operate at a specific value of the gain $g = 1/T$,
for a fixed value of the transmissivity $T^2$. This is the only choice that makes
the fidelity independent of $\beta$. In section \ref{SecNonunityGain} we adopt a
more general approach by letting $g$ be a fully free parameter and performing
appropriate optimization procedures. \\
In both cases, the maximization is carried out
at fixed (finite) squeezing $r$, and at fixed $\tau$, $n_{th}$, and $R$.
From an operational point of view, fixing these parameters is equivalent
to assume control on the characteristics of the experimental apparatus,
including the inefficiency of the photo-detectors and the length and damping
rate of the noisy channel. Finally, concerning the experimental realization
of the two-mode non-Gaussian resources, Eqs.~(\ref{squeezBell}) and (\ref{squeezCat}),
a detailed theoretical proposal is put forward in Ref.~\cite{CVTelepNoi}. The experimental
realization of the single-mode version of the state $|\psi\rangle_{SC}$ has been
reported in Ref.~\cite{GrangierCats}.

\section{$\beta$-independent optimal fidelity}
\label{SecUnityGain}
\begin{figure}[h]
\centering
\includegraphics*[width=9cm]{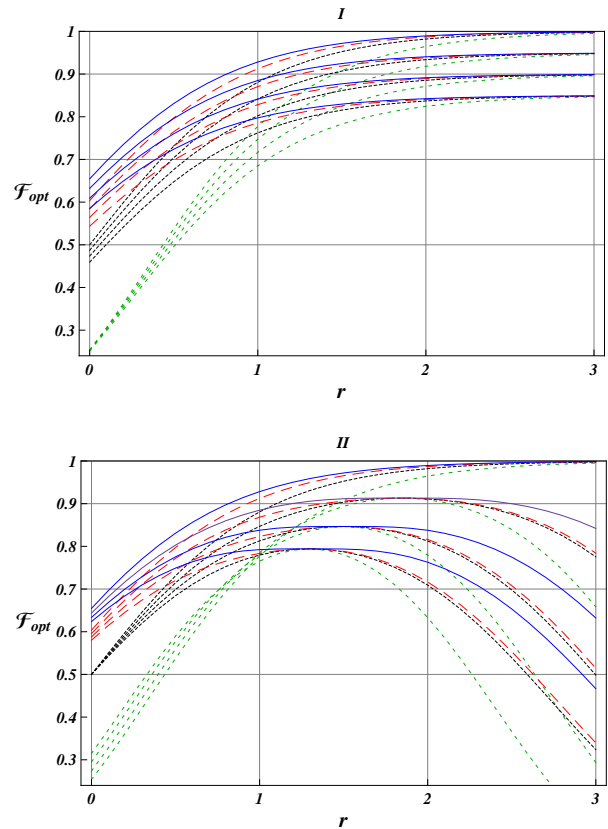}
\caption{(Color online) Optimal fidelities of teleportation $\mathcal{F}_{opt}$, as functions of the squeezing parameter $r$, for different values of the parameters $\tau$, $n_{th}$, and $R$. The fidelities correspond to the teleportation of single-mode input coherent states $|\beta\rangle$ using two-mode squeezed Bell-like states (full line) or two-mode squeezed cat-like states (dashed line) as entangled resources.
The fidelities associated with the two-mode squeezed vacuum (dotted line) and to the Buridan donkey states (long-dotted line) are reported for comparison. In panel I, $\tau=0$, $n_{th}=0$, and the reflectivity
is fixed at the values $R^{2}=0,\,0.05,\,0.1,\,0.15$. For each entangled resource (associated with a specific plot style), the corresponding curves are ordered from top to bottom with increasing $R^{2}$.
In panel II, $n_{th}=0$, $R=0$, and the reduced time is fixed at the values $\tau=0,\,0.1,\,0.2,\,0.3$.
For each entangled resource (associated to a specific plot style) the corresponding curves are ordered from top to bottom with increasing $\tau$.}
\label{FigTelFigGain1}
\end{figure}

In this section, we analyze the success probability of quantum teleportation
for the gain $g$ fixed to be $g \, = \, 1/T$. It is immediate to verify that
with this choice, the fidelity becomes $\beta$-independent (see appendix \ref{AppendixFid}).
This choice allows to assume no knowledge about the alphabet of input coherent states,
while in the next section we will assume a partial knowledge on the input states over an
interval of values of $\beta$. For $g\,=\, 1/T$, the expressions for the fidelities
$\mathcal{F}_{SB}(r,\delta)$, $\mathcal{F}_{SC}(r,\delta,|\gamma|)$,
and $\mathcal{F}_{SB2}(r,\delta)$
(where the superscript $(g)$ has been removed) greatly simplify.
At fixed $r$, $\tau$, $n_{th}$, and $R$, the optimal fidelities of teleportation
are defined as:
\begin{eqnarray}
&& \mathcal{F}_{opt}^{(SB)} \,=\, \max_{\delta} \, \mathcal{F}_{SB}(r,\delta)  \,,
\label{FidSBoptg1} \\
&& \mathcal{F}_{opt}^{(SB2)} \,=\, \max_{\delta} \, \mathcal{F}_{SB2}(r,\delta)  \,,
\label{FidSB2optg1} \\
&& \mathcal{F}_{opt}^{(SC)} \,=\, \max_{\delta,|\gamma|} \, \mathcal{F}_{SC}(r,\delta,|\gamma|)  \,.
\label{FidSCoptg1}
\end{eqnarray}
In Fig.~\ref{FigTelFigGain1} we plot the optimal fidelities,
corresponding to the three classes of non-Gaussian resources,
as functions of the squeezing $r$ at different values of the
parameters $\tau$, $n_{th}$, and $R=\sqrt{1-T^2}$.
For comparison, the fidelity associated with the Gaussian
squeezed vacuum (twin beam) is reported as well.
In order to understand the separate effects of the two
different sources of decoherence on the degradation
of the fidelity, we consider two cases:
$(i)$ decoherence due to imperfect Bell measurements alone, i.e. $R>0$ and $\tau=0$
(see Fig.~\ref{FigTelFigGain1} panel I);
$(ii)$ decoherence due to propagation in noisy channels alone, i.e. $R=0$ and $\tau >0$
(see Fig.~\ref{FigTelFigGain1} panel II).

In the first case, the fidelities grow monotonically, with increasing $r$, tending towards an
asymptotic saturation value. This behavior is equivalent to that observed in the instance of
an ideal protocol with noisy resources \cite{CVTelepNoisyNoi}. Indeed, the case of a nonideal
teleportation protocol with noisy (mixed) resources is equivalent to the case of a nonideal
protocol with pure resources but with a larger amount of noise.

In the second case, as $r$ increases, the fidelity first increases up to
a $\tau$-dependent maximum $r_{max}(\tau)$, and then decreases for larger values of $r$.
This behavior can be explained observing that there are two competing
effects associated to increasing the degree of squeezing. The first effect
is constructive and is due to the enhanced affinity of the entangled resource
with an EPR state for increasing $r$. This constructive effect is contrasted
by a disruptive one due to the optical photons generated by the squeezing that
add to the thermal photons of the channel (initially set to zero). For
not too large values of $r$, the first effect dominates, until a maximum
is reached at $r = r_{max}(\tau)$. For $r > r_{max}(\tau)$ the disruptive
effect becomes dominant, and the increasingly large number of optical photons
amplifies the decoherence, leading to a strong suppression of the fidelity.
The interplay between squeezing $r$ and channel decay rate $\tau$ can be
understood quantitatively by investigating the structure of the output
characteristic function $\chi_{out}$ Eq.~(\ref{chioutfinale}) that enters
in the expression of the fidelity (\ref{Fidelitychi}).
For $g T = 1$, $\chi_{out}$ takes the form
\begin{eqnarray}
\chi_{out}(x_{2},p_{2})= &&e^{-\frac{1}{2}\Gamma_{\tau,R}(x_{2}^{2}+p_{2}^{2})} \; \chi_{in}(x_{2},p_{2}) \times \nonumber \\ && \nonumber \\
&&\chi_{res}\left(x_{2},-p_{2};e^{-\frac{\tau}{2}}x_{2},e^{-\frac{\tau}{2}}p_{2}\right) \nonumber  \,,
\end{eqnarray}
where $\Gamma_{\tau,R}$ is given by Eq.~(\ref{Gammadef}).
We see that, if $\tau \neq 0$, the exponential weights $e^{-\tau/2}$
introduce an asymmetry between the two modes of the resource in the expression of the
characteristic function $\chi_{res}$. This asymmetry is responsible for the decrease
of the fidelity for $r >r_{max}(\tau)$ at $\tau\neq 0$. The important ensuing conclusion
is that it is in fact detrimental to increase too much the squeezing when the
losses cannot be reduced strongly. Therefore, the primary experimental goal should
always be that of reducing the losses rather than incrementing the squeezing.

Moreover, Fig.~\ref{FigTelFigGain1} (Panel II) shows that indeed at current experimentally
attainable values of the squeezing, i.e. $r \lesssim 1.5$, the nonideal
teleportation protocol operates already in the regime of best efficiency,
and both the squeezed Bell-like resources (\ref{squeezBell})
and the squeezed cat-like resources (\ref{squeezCat}) perform
much better than the corresponding (i.e. at the same squeezing) Gaussian resources.
This result generalizes and confirms the analogous behavior observed in the instance of ideal
protocols \cite{CVTelepNoi,CVTelepNoisyNoi}. On the contrary, as already anticipated in the
previous section, the Buridan donkey resources allow for teleportation fidelities
even worse than those associated with the Gaussian twin beam. Indeed, from Eq.~(\ref{FidelitySqBell2})
it follows that the optimal $\beta$-independent fidelity $\mathcal{F}_{opt}^{(SB2)}$ is obtained (with $g=1/T$) by letting $\delta=0$. In this case, the Buridan donkey state trivially reduces to a two-mode squeezed Fock state.

It is worth noting that, at $g T = 1$ and any fixed value of $\tau$, all the non-Gaussian resources
share with the Gaussian one the same maximum value of the fidelity, obtained at the same
value $r_{max}(\tau)$ of the squeezing parameter. This implies that, at given $\tau$, one
can determine the value $r_{max}(\tau)$ for all the various resources by just considering the
simple Gaussian instance. A straightforward computation then yields
\begin{equation}
\exp\{2\;r_{max}\} = \sqrt{\frac{\cosh (\tau/2) + 1}{\cosh (\tau/2) - 1}} \,.
\end{equation}
We see that, for increasing $\tau$, the range $[0,r_{max}(\tau)]$ of best efficiency reduces.
\begin{figure}[h]
\centering
\includegraphics*[width=8.5cm]{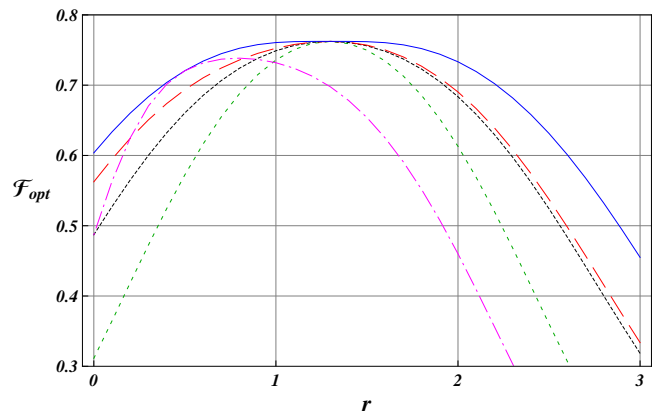}
\caption{(Color online) Optimal fidelities of teleportation $\mathcal{F}_{opt}$,
as functions of the squeezing parameter $r$,
with $\tau=0.3$, $n_{th}=0$, and $R^{2}=0.05$.
The different lines represent the fidelity of teleportation
of input coherent states $|\beta\rangle$, corresponding, respectively,
to the following entangled resources:
Squeezed Bell-like state(full line),
squeezed cat-like state (dashed line),
squeezed vacuum (dotted line),
Buridan donkey state (long-dotted line),
and photon-subtracted squeezed state (dot-dashed line).}
\label{FigTelFigGain1due}
\end{figure}

Finally, we consider the combined effect of the two decoherence mechanisms.
In Fig.~\ref{FigTelFigGain1due} we plot the optimal fidelities associated to
the different classes of non-Gaussian resources with the experimental parameters
fixed at the values $\tau=0.3$, $n_{th} = 0$, and $R^{2}=0.05$. In
this case, the simultaneous presence of the two effects leads to
a strong suppression of the fidelity both with respect to the ideal case and
to each of the two nonideal cases taken separately. A regime of best efficiency
is still present, but significantly reduced.

In Fig.~\ref{FigTelFigGain1due}, we also report the fidelity of teleportation
associated with the two-mode photon-subtracted squeezed states:
\begin{eqnarray}
&&|\psi\rangle_{PSS}= \mathcal{N} a_{1}a_{2}S_{12}(\zeta)|0,0\rangle \nonumber \\
&& \nonumber \\
&&=\mathcal{N}e^{i\phi} S_{12}(\zeta) \left\{-|0,0\rangle + e^{i\phi}\tanh r |1,1\rangle\right\} \,,
\label{PhotSubtractsqueez}
\end{eqnarray}
where $\mathcal{N}=(1+\tanh^{2}r)^{-1/2}$ is the normalization \cite{CVTelepNoi}.
This non-Gaussian resource belongs, as a particular subcase,
to the class of the squeezed Bell-like resources. Indeed,
Eq.~(\ref{squeezBell}) reduces to Eq.~(\ref{PhotSubtractsqueez}) for $\delta=\arctan (\tanh r)$
(with $\phi=\pi$ and $\theta=0$).
The interest in the resources (\ref{PhotSubtractsqueez}) is due to the fact that
such states have been already produced in the laboratory \cite{ExpdeGauss2,Grangier}.
Furthermore, the corresponding de-Gaussification scheme can be easily integrated
in the B-K teleportation protocol \cite{Opatrny}.
As in the ideal instance \cite{CVTelepNoi}, also in the realistic case,
see Fig.~\ref{FigTelFigGain1due},
the performance of the non-Gaussian resource (\ref{PhotSubtractsqueez})
is intermediate between that of the Gaussian twin beam and of the
squeezed Bell-like states for $0\leq r \lesssim 1$,
while, for $r \gtrsim 1$, it degrades much faster.
The affinity to the two-mode squeezed vacuum of
state (\ref{PhotSubtractsqueez}) decreases for growing $r$;
correspondingly, the resource becomes more and more non-Gaussian.
Moreover, for some intervals of values of the squeezing parameter
the photon-subtracted squeezed states behave better than the
squeezed cat-like states. It is worth noticing that there exists
a "crossing" value of $r$ at which the optimal squeezed Bell-like
state reduces to the photon-subtracted squeezed state and thus the
corresponding fidelities of teleportation coincide,
see Fig.~\ref{FigTelFigGain1due}.

In this section we have considered always the case $g T = 1$.
The scenario changes dramatically if $g T \neq 1$. Indeed, in this case
the analytical expressions of the fidelities (\ref{FidelitySqBell}),
(\ref{FidelitySqCat}), (\ref{FidelityTwB}), (\ref{FidelitySqBell2}),
depend on the coherent amplitude $\beta$. This dependence
affects quite significantly the behavior of the fidelities as we will
see in the next section.

\section{Average optimal fidelity and one-shot fidelity}
\label{SecNonunityGain}
In order to investigate possible improvements in the efficiency of the teleportation protocol,
in this section we aim at optimizing the success probability of teleportation
assuming $g$ as a further free optimization parameter.
The fidelity of teleportation as a function of the gain is studied
for several input states in Ref.~\cite{TelepIde}, while displacement strategies
are considered in Ref.~\cite{TelepTailored}, in order to improve the
output quality for a reduced alphabet of possible input states.
These two important works are concerned with the study of the ideal protocol
implemented using Gaussian resources. The effect of absorption due to propagation
in fibers is studied in Ref.~\cite{TelepChizhov}, where, for the case of Gaussian
resources, it is shown that the gain-optimized fidelity of teleportation
is strongly suppressed. \\
Let us now describe the optimization procedure applied to the instance of non-Gaussian resources.
Following the approach of Refs.~\cite{TelepChizhov,TelepTailored},
we define the average fidelity $\mathcal{\overline{F}}$ by averaging
the $\beta$-dependent fidelity $\mathcal{F}(\beta)$ over the
set of input coherent states $|\beta\rangle$ as follows:
\begin{eqnarray}
&&\mathcal{\overline{F}} \,=\, \int d^{2}\beta \; \mathcal{F}(\beta) \,
p(\beta) \,,
\label{averfid} \\
&& \nonumber \\
&&p(\beta)=(\pi \sigma)^{-1}\exp\{-\sigma^{-1} \, |\beta|^{2}\} \,,
\label{weight}
\end{eqnarray}
where the function $p(\beta)$ (\ref{weight}) is a Gaussian distribution centered at $\beta=0$.
The variance parameter $\sigma$ determines the cutoff of the amplitude $\beta$,
and thus the reduced alphabet that one considers. We will compare our results
with the quantum benchmark for the storage and transmission of coherent states
distributed according to Eq.~(\ref{weight}) \cite{Benchmark}. This benchmark
is equivalent to the upper bound $\mathcal{F}_{class}$ achievable with any
classical strategy, and satisfying the inequality \cite{Benchmark}:
\begin{equation}
\mathcal{F}_{class} \,\leq\, \frac{\sigma+1}{2\sigma+1} \,.
\label{BenchmarkCoh}
\end{equation}
The first step of our optimization procedure, employing Eqs.~(\ref{FidelitySqBell}) and (\ref{FidelitySqCat}), is to determine the average fidelities $\mathcal{\overline{F}}_{SB}^{(g)}(r,\delta)$
and $\mathcal{\overline{F}}_{SC}^{(g)}(r,\delta,|\gamma|)$, whose expressions
we do not report because their long and cumbersome structure is not particularly illuminating.
Further, we do not apply the optimization procedure to the teleportation fidelity associated to
Buridan donkey resource as we have already showed that no enhancement can be obtained compared
to the Gaussian twin beam resource.
At fixed squeezing $r$ and experimental parameters $\tau$, $n_{th}$, and $R$,
we define the optimal values  $g_{opt}$, $\delta_{opt}$, and $|\gamma_{opt}|$ of the free parameters
as those that maximize the fidelities averaged over the values of $\beta$ weighted according
to the normal distribution $p(\beta)$ (\ref{weight}):
\begin{equation}
\mathcal{\overline{F}}_{SB}^{(g_{opt})}(r,\delta_{opt}) \doteq \max_{\{g,\delta\}} \mathcal{\overline{F}}_{SB}^{(g)}(r,\delta)  ,
\label{optimparamSB}
\end{equation}
\begin{equation}
\mathcal{\overline{F}}_{SC}^{(g_{opt})}(r,\delta_{opt},|\gamma_{opt}|) \doteq \max_{\{g,\delta,|\gamma|\}} \mathcal{\overline{F}}_{SC}^{(g)}(r,\delta,|\gamma|) .
\label{optimparamSC}
\end{equation}
The optimal values of the parameters are determined numerically.
Next, we introduce the one-shot fidelities $\mathcal{F}_{1s}$
as the non-averaged fidelities evaluated at the optimal values
of the parameters and a fixed value of $\beta$:
\begin{equation}
\mathcal{F}_{1s}^{(SB)}(\beta,r) \, \doteq \, \mathcal{F}_{SB}^{(g_{opt})}(\beta,r,\delta_{opt}) \,,
\label{avoptFidSB}
\end{equation}
\begin{equation}
\mathcal{F}_{1s}^{(SC)}(\beta,r) \, \doteq \, \mathcal{F}_{SC}^{(g_{opt})}(\beta,r,\delta_{opt},|\gamma_{opt}|) \,.
\label{avoptFidSC}
\end{equation}
For each possible value of $\beta$ one can estimate the success probability
of teleportation associated with a specific event.
The functions $\mathcal{F}_{1s}(\beta,r)$ yield the teleportation fidelities
at given squeezing $r$ and for an input coherent state with specific amplitude $\beta$.
Partial information about the alphabet of input states, quantified by the choice of the
variance $\sigma$ in the distribution (\ref{weight}), can be exploited to obtain a refinement
of the optimization procedure.
Indeed, we expect that smaller values of $\sigma$, corresponding to a better knowledge
of the alphabet of input states, will lead to higher values of the one-shot fidelities.

\subsection{Fidelities: Variable $r$, fixed $\tau$}
\begin{figure}[h]
\centering
\includegraphics*[width=9cm]{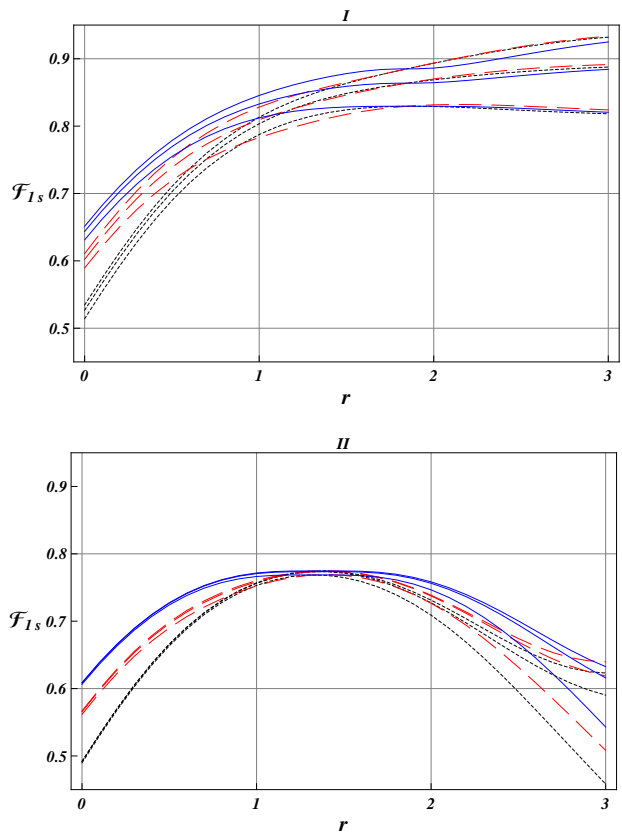}
\caption{(Color online) One-shot fidelity of teleportation $\mathcal{F}_{1s}$,
for input coherent states $|\beta\rangle$,
as a function of the squeezing parameter $r$,
with $\tau=0.3$, $n_{th}=0$, $R^{2}=0.05$,
for the following resources: Squeezed Bell-like state (full line);
squeezed cat-like state (dashed line); and a squeezed vacuum (dotted line).
In panel I $\beta=1,\,2,\,3$ and $\sigma=10$. In panel II $\beta=3,\,5,\,10$
and $\sigma=100$. For each entangled resource (associated with a specific plot style),
the corresponding curves are ordered from top to bottom with increasing $\beta$.}
\label{FigTelnonunityGain}
\end{figure}

In Fig.~\ref{FigTelnonunityGain}, at the same fixed parameters of Fig.~\ref{FigTelFigGain1due},
$\tau=0.3$, $n_{th}\approx 0$, $R^{2}=0.05$,
we plot the one-shot fidelity $\mathcal{F}_{1s}$ as a function of $r$,
both for the non-Gaussian resources and for the Gaussian twin beam.
In panel I we plot the one-shot fidelities for $\sigma=10$ and
$\beta=1,2,3$; in panel II we plot the one-shot fidelities for  $\sigma=100$ and
$\beta=3,5,10$ (obviously $|\beta|^2$ must fall in the interval $[0,\sigma]$).
Let us notice that according to Eq.~(\ref{BenchmarkCoh}),
the quantum benchmarks for the two choices of $\sigma$ are:
$\mathcal{F}_{class}(\sigma=10) \approx 0.523$ and
$\mathcal{F}_{class}(\sigma=100) \approx 0.502$.

Panel I of Fig.~\ref{FigTelnonunityGain}, corresponding to $\sigma=10$
and small values of $|\beta|$, shows that, as soon as $r$ is different from zero,
all the resources yield fidelities above the quantum benchmark.
Moreover, one observes a significant enhancement of the fidelities with respect
to the $\beta$-independent ones of Fig.~\ref{FigTelFigGain1due} (corresponding
to $g = 1/T$). Indeed, while the $\beta$-independent fidelities in Fig.~\ref{FigTelFigGain1due}
are well below the value $0.8$, all the one-shot fidelities of panel I of Fig.~\ref{FigTelnonunityGain}
lie above this value for squeezing $r$ ranging from about $0.8$ to about $1.2$, depending on the resource
being considered. Also in this case the non-Gaussian resources exhibit better performances with respect to the Gaussian ones.
Panel II of Fig.~\ref{FigTelnonunityGain} shows that for $\sigma=100$ and larger values
of $|\beta|$, the enhancement of the fidelity is quite modest compared to
the $\beta$-independent fidelity, reported in Fig.~\ref{FigTelFigGain1due}.
This result is not surprising because a variance $\sigma=100$ obviously
allows less knowledge on the alphabet of input states with respect to the case $\sigma=10$.
It is important to remark that the curves corresponding to the same entangled resource
but different values of $\beta$ become effectively distinguishable only $r \gtrsim 1$.

\subsection{Fidelities: Variable $\tau$, fixed $r$}
\begin{figure}[h]
\centering
\includegraphics*[width=9cm]{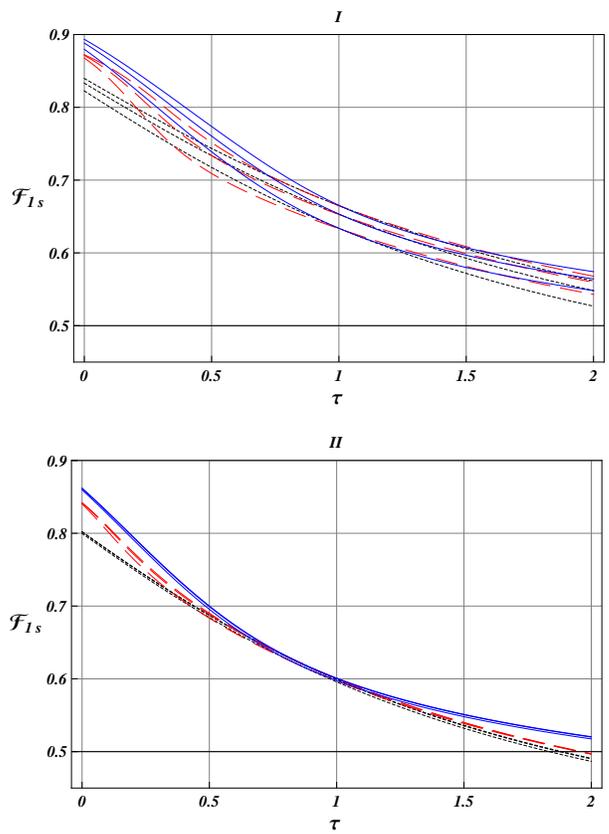}
\caption{(Color online) One-shot fidelity of teleportation $\mathcal{F}_{1s}$,
for input coherent states $|\beta\rangle$,
as a function of the reduced time $\tau$,
with $r=0.8$, $n_{th}=0$, $R^{2}=0.05$, for
the following entangled resources:
Squeezed Bell-like state (full line);
squeezed cat-like state (dashed line);
and squeezed vacuum (dotted line).
In panel I $\beta=1,\,2,\,3$ and $\sigma=10$.
In panel II $\beta=3,\,5,\,10$ and $\sigma=100$.
For each entangled resource (associated to a specific plot style),
the corresponding curves are ordered from top to bottom
with increasing $\beta$.}
\label{FigTelnonunityGain2}
\end{figure}

We now study the fidelities as functions of the reduced time,
or effective length of the fiber $\tau$, at fixed squeezing $r$.
To this aim, fixing the squeezing parameter at the intermediate value $r=0.8$,
with $n_{th} = 0$, $R^{2}=0.05$, we investigate the behavior of the one-shot fidelities.
In Fig.~\ref{FigTelnonunityGain2}, choosing the same values of $\beta$ and $\sigma$
as in Fig.~\ref{FigTelnonunityGain}, we plot $\mathcal{F}_{1s}$ as a function of $\tau$.

Fig.~\ref{FigTelnonunityGain2} shows that the teleportation fidelity
remains above the classical threshold up to significantly large values of $\tau$.
At $\sigma=10$ and for small values of of the coherent amplitude $\beta$ (see panel I),
the one-shot fidelities associated to the same resource
but to different values of $\beta$ are distinguishable. Viceversa,
at $\sigma=100$ and for larger values of $\beta$ (see panel II),
the fidelities associated to the same resource but different values
of $\beta$ are virtually indistinguishable, and thus effectively
$\beta$-independent. Comparing the performance of the different
resources in the time domain, one can distinguish three regimes:
For short and long times the one-shot fidelities associated to non-Gaussian
resources exhibit a significant enhancement compared to the Gaussian instance,
while for intermediate times there is a substantial equivalence in the
performance of Gaussian and non-Gaussian resources.

\section{Conclusions}
\label{secConclusions}
In this paper we have investigated the performance of non-Gaussian entangled resources
in nonideal protocols of quantum teleportation of input coherent states.
We have resorted to the characteristic function formalism
for the description of protocols affected by decoherence.
We have discussed how the effects of decoherence stemming from
photon losses in fiber and imperfect Bell measurements
affect the success probability of teleportation. In particular,
we have established that, while the fidelities associated to
different resources remain above the classical benchmark for
quite long times, the non-Gaussian resources perform always better
than the Gaussian ones, in the ideal as well as in the nonideal
quantum teleportation protocol.

The present analysis should be extended to include teleportation
of two-mode states using multimode non-Gaussian resources  \cite{twomodeinputTelep}.
It would be also interesting to consider the optimization of teleportation protocols
with non-Gaussian entangled resources with respect to local properties (such as
single-mode squeezing), by extending the existing schemes for the local optimization
of Gaussian resources \cite{GaussianOptimization}.

\acknowledgments
This work has been realized in the framework of the FP7 STREP Project
HIP (Hybrid Information Processing). We acknowledge financial support also from
MIUR under FARB Funds 2007 and 2008, from INFN under Iniziativa Specifica
PG62, and CNR-INFM Coherentia. F. I. acknowledges financial support from
the ISI Foundation for Scientific Interchange.

\appendix
\section{Realistic teleportation protocol in the characteristic function representation}
\label{AppendixRealTel}
Here we outline in some details the description of the nonideal Braunstein-Kimble
teleportation protocol in terms of the characteristic function formalism.
Let us consider the Bell measurement Fig.~\ref{FigRealQuantTel2}.
The $50/50$ beam splitter $BS_{1}$ acts on the initial density operator $\rho_{0}$
through a $SU(2)$ transformation $U_{BS_{1}}(\frac{\pi}{4})$,
consisting in a balanced mixing of the modes $in$ and $1$:
\begin{equation}
\rho_{0}'=U_{BS_{1}}\left(\frac{\pi}{4}\right)\, \rho_{0}\, U_{BS_{1}}^{\dag}\left(\frac{\pi}{4}\right) \,.
\label{UBS1}
\end{equation}
This transformation yields the following relation among the phase-space variables
corresponding to the input and output modes of the beam splitter:
\begin{equation}
\left\{\begin{array}{rrr}
  x_{in}'=\frac{1}{\sqrt{2}}(x_{in}+x_{1}) \;, & p_{in}'=\frac{1}{\sqrt{2}}(p_{in}+p_{1}) \;,
  \\
  & \\
  x_{1}'=\frac{1}{\sqrt{2}}(x_{in}-x_{1}) \;, &
  p_{1}'=\frac{1}{\sqrt{2}}(p_{in}-p_{1}) \, ,
\end{array} \right.
\label{BStransformXP}
\end{equation}
where the canonical variables $x_{j}'$ and $p_{j}'$
correspond to the primed modes $j'$ $(j'= in'\,,1')$.
Using Eq.~(\ref{BStransformXP}), the characteristic function $\chi_{0}'$
associated to the density operator $\rho_{0}'$ reads:
\begin{eqnarray}
&&\chi_{0}'(x_{in}',p_{in}';x_{1}',p_{1}';x_{2},p_{2}) \,=\,
\nonumber \\
&& \nonumber \\
&&\chi_{in}\left(\frac{1}{\sqrt{2}}(x_{in}'+x_{1}')\,,\frac{1}{\sqrt{2}}(p_{in}'+p_{1}')\right)\; \times
\nonumber \\
&& \nonumber \\
&& \chi_{res}\left(\frac{1}{\sqrt{2}}(x_{in}'-x_{1}')\,,\frac{1}{\sqrt{2}}(p_{in}'-p_{1}')\,;x_{2}\,,p_{2}\right)
.
\label{charfuncBS1}
\end{eqnarray}
The fields $1'$ and $in'$ at the output of the beam splitter $BS_{1}$
feed the input ports of the two beam splitters
$BS_{2}$ and $BS_{3}$, respectively, with equal transmissivity $\eta$,
see Fig.~\ref{FigRealQuantTel2}.
The remaining input ports of the beam splitters are fed with the field modes $3'$ and $4'$,
initially in the vacuum states $|0\rangle_{4'}$ and $|0\rangle_{3'}$,
corresponding to the characteristic functions
\begin{equation}
\chi_{k}(x_{k}',p_{k}') \,=\, e^{-\frac{1}{4}(x_{k}^{'2}+p_{k}^{'2})}
\,, \quad  k=3,4 \,.
\label{chivacua}
\end{equation}
The beam splitters $BS_{2}$ and $BS_{3}$,
associated to the $SU(2)$ transformations $U_{BS_{j}}(\eta)$ $(j=2,3)$,
transform the variables $x_{j}'$ and $p_{j}'$ $(j=in,1,3,4)$ according to
the following relations:
\begin{equation}
\left\{\begin{array}{rrr}
  x_{1}''=T x_{1}'-Rx_{4}' \;, & p_{1}''=T p_{1}'-Rp_{4}' \;,
  \\
  & \\
  x_{4}''=T x_{4}'+Rx_{1}' \;, & p_{4}''=T p_{4}'+Rp_{1}' \;,
\end{array} \right.
\label{BS2transf} \\
\end{equation}

\begin{equation}
\left\{\begin{array}{rrr}
  x_{in}''=T x_{in}'-Rx_{3}' \;, & p_{in}''=T p_{in}'-Rp_{3}' \;,
  \\
  & \\
  x_{3}''=T x_{3}'+Rx_{in}' \;, & p_{3}''=T p_{3}'+Rp_{in}' \;,
\end{array} \right.
\label{BS3transf}
\end{equation}
where $T^2=\eta$ and $R^2=1-\eta$ denote, respectively,
transmissivity and reflectivity of the beam splitters,
and double primes denote output modes.
At the output of $BS_{2}$ and $BS_{3}$, the transformed five-mode density operator reads
\begin{eqnarray}
\rho_{BS}= && U_{BS_{3}}(\eta)U_{BS_{2}}(\eta)\rho_{0}'\otimes |0\rangle_{3'}\,_{3'}\langle 0|\otimes
\nonumber \\
&& \nonumber \\
&&|0\rangle_{4'}\,_{4'}\langle 0|U_{BS_{2}}^{\dag}(\eta)U_{BS_{3}}^{\dag}(\eta) \,,
\label{rhoBS}
\end{eqnarray}
where $\rho_{0}'$ is given by Eq.~(\ref{UBS1}).
By exploiting the above transformations, a simple formal manipulation yields
the characteristic function $\chi_{BS}$ corresponding to $\rho_{BS}$:
\begin{widetext}
\begin{eqnarray}
&&\chi_{BS}(x_{in}'',p_{in}'';x_{1}'',p_{1}'';x_{2},p_{2};x_{3}'',p_{3}'';x_{4}'',p_{4}'')= \chi_{in}\left(\frac{1}{\sqrt{2}}\left[T x_{in}''+R x_{3}''+T x_{1}''+R x_{4}''\right]\,,\frac{1}{\sqrt{2}}\left[T p_{in}''+R p_{3}''+T p_{1}''+R p_{4}''\right] \right) \times \nonumber \\
&& \nonumber \\
&&\chi_{res}\left(\frac{1}{\sqrt{2}}\left[T x_{in}''+R x_{3}''-T x_{1}''-R x_{4}''\right]\,,\frac{1}{\sqrt{2}}\left[T p_{in}''+R p_{3}''-T p_{1}''-R p_{4}''\right];x_{2},p_{2}\right) \chi_{3}\Big(T x_{3}''-R x_{in}''\,,T p_{3}''-R p_{in}'' \Big)
\times \nonumber \\
&& \nonumber \\
&&\chi_{4}\Big(T x_{4}''-R x_{1}''\,,T p_{4}''-R p_{1}'' \Big) \,,
\label{chiBellMeas1}
\end{eqnarray}
\end{widetext}
where the functions $\chi_{k}$, with $k=3,4$, are given by Eq.~(\ref{chivacua}).
The Bell measurement, that consist in the two homodyne measurements of the variables
$p_{in}''$ and $x_{1}''$, yields the results $p_{in}''=\tilde{p}$ and $x_{1}''=\tilde{x}$,
and transforms the characteristic function $\chi_{BS}$ to (see appendix \ref{AppendixHomoMeas}
for details):
\begin{eqnarray}
&&\chi_{Bm}(x_{2},p_{2}) = \frac{\mathcal{P}^{-1}(\tilde{p},\tilde{x})}{(2\pi)^{2}}
\int dx_{in}''dp_{1}'' \;  e^{i x_{in}''\tilde{p}-i \tilde{x} p_{1}''} \times \nonumber \\
&& \nonumber \\
&& \chi_{BS}(x_{in}'',0;0,p_{1}'';x_{2},p_{2};0,0;0,0) \, .
\label{chiBellMeas2}
\end{eqnarray}
Here $\mathcal{P}(\tilde{p},\tilde{x})$ is the distribution function of the outcomes
$\tilde{p}$ and $\tilde{x}$. It reads:
\begin{eqnarray}
&&\mathcal{P}(\tilde{p},\tilde{x}) \, =  \,
Tr[|\tilde{p}\rangle_{in''}\,_{in''}\langle\tilde{p}| \otimes
|\tilde{x}\rangle_{1''}\,_{1''}\langle\tilde{x}| \; \rho_{BS}] \nonumber \\
&& \nonumber \\
&&=\frac{1}{(2\pi)^{2}}\int dx_{in}''dp_{1}''  \;
\exp\{i x_{in}''\tilde{p}-i \tilde{x} p_{1}''\}\times \nonumber \\
&& \nonumber \\
&&  \chi_{BS}(x_{in}'',0;0,p_{1}'';0,0;0,0;0,0) \,.
\label{xpdistribfunction}
\end{eqnarray}
Let us notice that in Eq.~(\ref{chiBellMeas2}) the modes $3''$ and $4''$,
associated with the unused output ports of the two fictitious beam splitters,
have been traced out. This is equivalent to setting to zero in the function $\chi_{BS}$
the variables $x_{j}''$ and $p_{j}''$ with $j=3,4$. Meanwhile, mode $2$
propagates in a noisy channel towards Bob's location.
The dynamics of this mode is described by the master equation (\ref{MasterEq}),
and the corresponding diffusion equation for the characteristic function
$\chi_{t}(x_{2},p_{2})$ is \cite{WallsMilburn,DecohReview}:
\begin{eqnarray}
&&\partial_{t} \; \chi_{t}(x_{2},p_{2}) \,=\, -
\frac{\Upsilon}{2} \Big[ \Big(\frac{1}{2}+n_{th} \Big)(x_{2}^{2}+p_{2}^{2})  \nonumber \\
&& \nonumber \\
&&  +x_{2}\partial_{x_{2}}+y_{2}\partial_{y_{2}}
 \Big] \; \chi_{t}(x_{2},p_{2}) \,.
\label{DiffusionEq}
\end{eqnarray}
Given the initial characteristic function $\chi_{Bm}(x_{2},p_{2})$ Eq.~(\ref{chiBellMeas2}),
the global characteristic function $\chi_{t}(x_{2},p_{2})$ following the Bell
measurement is given by the solution of Eq.~(\ref{DiffusionEq}):
\begin{eqnarray}
\chi_{t}(x_{2},p_{2}) \,=\, \chi_{Bm}(e^{-\frac{1}{2}\Upsilon t}x_{2},e^{-\frac{1}{2}\Upsilon t}p_{2}) \times
\nonumber \\
\nonumber \\
\exp\left\{-\frac{1}{2} (1-e^{-\Upsilon t})\left(\frac{1}{2}+n_{th}\right)(x_{2}^{2}+p_{2}^{2})\right\}.
\label{solchit}
\end{eqnarray}
After recovering the classical information, Bob performs on mode $2$
the displacement $\lambda= g(\tilde{x}+i \tilde{p})$, where $g$ is
the gain factor. One easily verifies that, given a state $\rho_{t}$
described by the characteristic function $\chi_{t}(x_{2},p_{2})$,
the displaced state $D_{2}(\lambda)\rho_{t}D_{2}^{\dag}(\lambda)$
is described by the characteristic function
\begin{equation}
\chi_{D}(x_{2},p_{2}) \,=\, \chi_{t}(x_{2},p_{2})\,  e^{i \sqrt{2}Re[\lambda] p_{2}-i \sqrt{2}Im[\lambda] x_{2}} \,.
\label{chidisplaced}
\end{equation}
In order to obtain the density operator $\rho_{out}$
(or, equivalently, the characteristic function $\chi_{out}(x_{2},p_{2})$),
of the final output state of the teleportation protocol, one must
take the average on all the possible outcomes $\tilde{x}$, $\tilde{p}$
of the Bell measurement:
\begin{eqnarray}
&&\chi_{out}(x_{2},p_{2}) =\int d\tilde{x} d\tilde{p} \;\, \mathcal{P}(\tilde{p},\tilde{x})
\; \chi_{D}(x_{2},p_{2}) \nonumber \\
&& \nonumber \\
&&= \frac{1}{(2\pi)^{2}} \int\, d\tilde{x}\, d\tilde{p}\, dx_{in}''\,dp_{1}'' \times \nonumber \\
&& \nonumber \\
&& \exp\left\{i \tilde{x} \left(\sqrt{2}g p_{2}- p_{1}''\right)-i \tilde{p}\left( \sqrt{2}g x_{2}-x_{in}'' \right)\right\} \times \nonumber \\
&& \nonumber \\
&&  \, \chi_{BS}(x_{in}'',0;0,p_{1}'';e^{-\frac{1}{2}\Upsilon t} x_{2},e^{-\frac{1}{2}\Upsilon t}p_{2};0,0;0,0)  \times \nonumber \\
&& \nonumber \\
&&  \exp\left\{-\frac{1}{2} (1-e^{-\Upsilon t})\left(\frac{1}{2}+n_{th}\right)(x_{2}^{2}+p_{2}^{2})\right\} \,.
\label{chiout1}
\end{eqnarray}
Finally, by exploiting the relation $(2\pi)^{-1}\int dk \exp\{i k x\} = \delta(x)$,
and recalling the expression Eq.~(\ref{chiBellMeas1}) of $\chi_{BS}$, we find that
Eq.~(\ref{chiout1}) reduces to Eq.~(\ref{chioutfinale}).

\section{Homodyne measurement in the characteristic function representation}
\label{AppendixHomoMeas}
In this appendix we describe the homodyne measurement for a multimode state
in the characteristic function representation.
In particular, we determine the expression for the density operator
and for its characteristic function of the reduced state after post-selection,
i.e. of the output state resulting from the homodyne measurements on
appropriately selected modes. Let us consider a three-mode quantum
state $\rho_{123}$, with characteristic function
$\chi_{123}(\alpha_{1},\alpha_{2},\alpha_{3}) \,=\, Tr[\rho_{123} \,
D_{1}(\alpha_{1})D_{2}(\alpha_{2})D_{3}(\alpha_{3})] \,\equiv \,
\chi_{123}(x_{1},p_{1};x_{2},p_{2};x_{3},p_{3})$,
with $\alpha_{j}=2^{-1/2}(x_{j}+i p_{j})$ $(j=1,2,3)$.
Let us recall that $\rho_{123}$ can be written in terms of the Weyl expansion:
\begin{eqnarray}
&&\rho_{123} \,=\, \frac{1}{\pi^{3}} \int d^{2}\alpha_{1}d^{2}\alpha_{2}d^{2}\alpha_{3} \;
\chi_{123}(\alpha_{1},\alpha_{2},\alpha_{3}) \times \nonumber \\
&& \nonumber \\
&&D_{1}(-\alpha_{1})D_{2}(-\alpha_{2})D_{3}(-\alpha_{3}) \,.
\label{Weylexp123}
\end{eqnarray}
Homodyne measurements performed on two modes, acting as two conditional measurements,
reduce the three-mode state to a single-mode one. Let $p_{1}=\tilde{p}$ and $x_{2}=\tilde{x}$
be the results of the homodyne measurements of the quadratures $p_{1}$ and $x_{2}$.
The homodyne measurements are expressed by projections on the quadrature eigenstates
$|\tilde{p}\rangle_{1}$ and $|\tilde{x}\rangle_{2}$. Thus the output state $\tilde{\rho}_{3}$
subsequent to the conditional measurements is:
\begin{equation}
\tilde{\rho}_{3} = \mathcal{P}^{-1}(\tilde{p},\tilde{x}) \, Tr_{12}[|\tilde{p}\rangle_{1}\,_{1}\langle\tilde{p}| \otimes |\tilde{x}\rangle_{2}\,_{2}\langle\tilde{x}| \; \rho_{123}] \,.
\label{Weylexp3}
\end{equation}
Here the normalization factor $\mathcal{P}(\tilde{p},\tilde{x})= Tr_{3}[\tilde{\rho}_{3}]$
represents the distribution function of the outcomes $\tilde{p}$ and $\tilde{x}$.
By using the relations
\begin{eqnarray}
&&\,_{1}\langle \tilde{p}|D_{1}(-\alpha_{1}) |\tilde{p}\rangle_{1} \,=\,
e^{-\frac{i}{2}x_{1}p_{1}+i x_{1}\tilde{p}} \delta(p_{1}) \,, \\
&& \nonumber \\
&&\,_{2}\langle \tilde{x}|D_{2}(-\alpha_{2}) |\tilde{x}\rangle_{2} \,=\,
e^{-\frac{i}{2}x_{2}p_{2}-i \tilde{x}p_{2}} \delta(x_{2}) \,,
\end{eqnarray}
Eq.~(\ref{Weylexp3}) writes
\begin{eqnarray}
&&\tilde{\rho}_{3} = \frac{\mathcal{P}^{-1}(\tilde{p},\tilde{x})}{(2\pi)^{3}} \int dx_{1}dp_{2}dx_{3}dp_{3} \; e^{i x_{1}\tilde{p}-i \tilde{x}p_{2}} \times \nonumber \\
&&\chi_{123}(x_{1},0;0,p_{2};x_{3},p_{3}) \,
D_{3}\left(-\frac{x_{3}+ip_{3}}{\sqrt{2}}\right) \,.
\label{Weylexp3bis}
\end{eqnarray}
Denoting by $\chi_{3}(x_{3},p_{3})$ the characteristic function of the state
$\tilde{\rho}_{3}$, we have
\begin{equation}
\tilde{\rho}_{3} \,=\, \frac{1}{2\pi} \int dx_{3}dp_{3} \; \chi_{3}(x_{3},p_{3})
D_{3}\left(-\frac{x_{3}+ip_{3}}{\sqrt{2}}\right) \,.
\label{Weylexp4}
\end{equation}
Comparing Eqs.~(\ref{Weylexp3bis}) and (\ref{Weylexp4}), we obtain
\begin{eqnarray}
&&\chi_{3}(x_{3},p_{3}) \,=\,  \frac{\mathcal{P}^{-1}(\tilde{p},\tilde{x})}{(2\pi)^{2}} \int dx_{1} dp_{2} \, \; e^{i x_{1}\tilde{p}-i \tilde{x}p_{2}} \times \nonumber \\
&& \nonumber \\
&&\chi_{123}(x_{1},0;0,p_{2};x_{3},p_{3})  \,.
\label{Weylexp5}
\end{eqnarray}
Moreover, it is easy to verify that
\begin{eqnarray}
&&\mathcal{P}(\tilde{p},\tilde{x}) \,=\,  \frac{1}{(2\pi)^{2}} \int dx_{1} dp_{2} \;\,
e^{i x_{1}\tilde{p}-i \tilde{x}p_{2} } \times \nonumber \\
&& \nonumber \\
&&\chi_{123}(x_{1},0;0,p_{2};0,0)  \,.
\label{Weylexp5Norm}
\end{eqnarray}

\section{Analytical expressions for the fidelities}
\label{AppendixFid}
In this appendix we report the analytical expressions for the fidelities
of teleportation of input coherent states using two classes of non-Gaussian
entangled resources: The squeezed Bell-like states (\ref{squeezBell}) and
the squeezed cat-like states (\ref{squeezCat}). For comparison we include
also the expressions of the fidelities associated to the Gaussian twin-beam
and of the Buridan donkey state (\ref{squeezBell2}.
The fidelity $\mathcal{F}_{SB}^{(g)}(r,\delta)$ for the squeezed Bell-like state
(\ref{squeezBell}) reads:
\begin{eqnarray}
&&\mathcal{F}_{SB}^{(g)}(r,\delta) \,=\,
\frac{4}{\Delta}e^{-\frac{4}{\Delta}(\tilde{g}-1)^{2}|\beta|^{2}}
\Big\{1+\frac{2e^{-4r-2\tau}}{\Delta^{4}}\times \nonumber \\
&& \nonumber \\
&&\Big((1+e^{\frac{\tau}{2}}\tilde{g})^{2}-e^{4r}(1-e^{\frac{\tau}{2}}\tilde{g})^{2}
\Big)^{2} \times
\nonumber \\
&& \nonumber \\
&&[\Delta^{2}-8\Delta(\tilde{g}-1)^{2}|\beta|^{2}+8(\tilde{g}-1)^{4}|\beta|^{4}]\sin^{2}\delta
\nonumber \\
&& \nonumber \\
&&+\frac{2e^{-2r-\tau}}{\Delta^{2}}[4(\tilde{g}-1)^{2}|\beta|^{2}-\Delta]\sin\delta
[\cos\delta \times \nonumber \\
&& \nonumber \\
&&\Big(-(1+e^{\frac{\tau}{2}}\tilde{g})^{2}+e^{4r}(1-e^{\frac{\tau}{2}}\tilde{g})^{2}\Big)
+\sin\delta \times \nonumber \\
&& \nonumber \\
&&\Big((1+e^{\frac{\tau}{2}}\tilde{g})^{2}+e^{4r}(1-e^{\frac{\tau}{2}}\tilde{g})^{2}\Big)] \Big\} \,,
\label{FidelitySqBell}
\end{eqnarray}
where
\begin{eqnarray}
\Delta \,\doteq \, e^{-2r-\tau}[(1+e^{\frac{\tau}{2}}\tilde{g})^{2}+e^{4r}(1-e^{\frac{\tau}{2}}\tilde{g})^{2} + \nonumber \\
\nonumber \\
2e^{2r+\tau}(1+\tilde{g}^{2}+2\Gamma_{\tau,R})],
\label{Deltadef}
\end{eqnarray}
\begin{equation}
\tilde{g} \,=\, g\, T \,,
\end{equation}
and $\Gamma_{\tau,R}$ is defined in Eq.~(\ref{Gammadef}).
The fidelity $\mathcal{F}_{SC}^{(g)}(r,\delta,|\gamma|)$
for the squeezed cat-like state (\ref{squeezCat}) reads:
\begin{eqnarray}
&&\mathcal{F}_{SC}^{(g)}(r,\delta,|\gamma|) \,=\,
\frac{4}{\Delta(1+e^{-|\gamma|^{2}}\sin 2\delta)}
\Big\{\cos^{2}\delta \times \nonumber \\
&& \nonumber \\
&&e^{-\frac{4}{\Delta}(\tilde{g}-1)^{2}|\beta|^{2}}
+e^{-|\gamma|^{2}-\frac{1}{\Delta}[(\tilde{g}-1)(\beta+\beta^{*})-e^{r}\tilde{g}|\gamma|]^{2}} \times
\nonumber \\
&& \nonumber \\
&&\sin \delta \cos\delta \, \Big( e^{\frac{1}{\Delta}[(\tilde{g}-1)(\beta-\beta^{*})+e^{-r}\tilde{g}|\gamma|]^{2}}+
c.c.\Big)
\nonumber \\
&& \nonumber \\
&& +\sin^{2}\delta \;
e^{-\frac{4}{\Delta}|(\tilde{g}-1)\beta-e^{r}|\gamma|(\tilde{g}-e^{\frac{\tau}{2}})|^{2}} \Big\}\,.
\label{FidelitySqCat}
\end{eqnarray}
For $\tilde{g} = 1$ and $\tau=R=n_{th}=0$, Eqs.~(\ref{FidelitySqBell}) and (\ref{FidelitySqCat})
reduce to the expressions of the fidelities associated to the ideal protocol with unit gain,
that had been originally determined in Ref.~\cite{CVTelepNoisyNoi}.
By letting $\delta=0$ in Eqs.~(\ref{FidelitySqBell}) and (\ref{FidelitySqCat}),
we recover the fidelity $\mathcal{F}_{TwB}^{(g)}(r)$ associated to the Gaussian
twin-beam resource:
\begin{equation}
\mathcal{F}_{TwB}^{(g)}(r) \,=\,
\frac{4}{\Delta}e^{-\frac{4}{\Delta}(\tilde{g}-1)^{2}|\beta|^{2}}\,.
\label{FidelityTwB}
\end{equation}
Let us notice that such a result can also be obtained by putting $|\gamma|=0$
in Eq.~(\ref{FidelitySqCat}).
For completeness, we also report the expression of the teleportation fidelity
associated to the Buridan donkey entangled resource (\ref{squeezBell2}):
\begin{eqnarray}
&&\mathcal{F}_{SB2}^{(g)}(r,\delta) \,=\,
\frac{4}{\Delta}e^{-\frac{4}{\Delta}(\tilde{g}-1)^{2}|\beta|^{2}}
\Big\{1+\frac{e^{-2r-\tau}}{\Delta^{2}}\times \nonumber \\
&& \nonumber \\
&&\Big[\Big((1+e^{\frac{\tau}{2}}\tilde{g})^{2}+e^{4r}(1-e^{\frac{\tau}{2}}\tilde{g})^{2}\Big) \times
\nonumber \\
&& \nonumber \\
&&[4(\tilde{g}-1)^{2}|\beta|^{2}-\Delta]+2\cos 2\delta \; e^{2r}(e^{\tau}\tilde{g}^{2}-1)\times
\nonumber \\
&& \nonumber \\
&&[\Delta-4(\tilde{g}-1)^{2}|\beta|^{2}]-2\sin2\delta \; (\tilde{g}-1)^{2}(\beta^{2}+\beta^{*2}) \times
\nonumber \\
&& \nonumber \\
&& \Big((1+e^{\frac{\tau}{2}}\tilde{g})^{2}-e^{4r}(1-e^{\frac{\tau}{2}}\tilde{g})^{2}\Big)
\Big]\Big\} \,. \nonumber \\
&&
\label{FidelitySqBell2}
\end{eqnarray}


\begin{thebibliography}{99}

\bibitem{PhysRep}
F. Dell'Anno, S. De Siena, and F. Illuminati, Phys. Rep. {\bf 428}, 53 (2006).

\bibitem{KimBS}
M. S. Kim, W. Son, V. Bu\v{z}ek, and P. L. Knight, Phys. Rev. A {\bf 65}, 032323 (2002).

\bibitem{KitagawaPhotsub}
A. Kitagawa, M. Takeoka, M. Sasaki, and A. Chefles, Phys. Rev. A {\bf 73}, 042310 (2006).

\bibitem{DodonovDisplnumb}
V. V. Dodonov and L. A. de Souza, J. Opt. B: Quantum Semiclass. Opt. {\bf 7}, S490 (2005).

\bibitem{Cerf}
N. J. Cerf, O. Kr\"uger, P. Navez, R. F. Werner, and M. M. Wolf, Phys. Rev. Lett. {\bf 95}, 070501 (2005).

\bibitem{QEstimNoi}
G. Adesso, F. Dell'Anno, S. De Siena, F. Illuminati, and L. A. M. Souza, Phys. Rev. A {\bf 79}, 040305 (2009).

\bibitem{CxKerrKorolkova}
T. Tyc and N. Korolkova, New J. Phys. {\bf 10}, 023041 (2008).

\bibitem{AgarTara}
G. S. Agarwal and K. Tara, Phys. Rev. A {\bf 43}, 492 (1991).

\bibitem{DeGauss1}
G. Bjork and Y. Yamamoto, Phys. Rev. A {\bf 37}, 4229 (1988).

\bibitem{DeGauss2}
Z. Zhang and H. Fan, Phys. Lett. A {\bf 165}, 14 (1992).

\bibitem{DeGauss3}
M. Dakna, T. Opatrn\'{y}, L. Kn\"{o}ll, and D. G. Welsch, Phys. Rev. A {\bf 55}, 3184 (1997).

\bibitem{DeGauss4}
M. S. Kim, E. Park, P. L. Knight, and H. Jeong, Phys. Rev. A {\bf 71}, 043805 (2005).

\bibitem{DeGauss5}
D. Menzies and R. Filip, Phys. Rev. A {\bf 79}, 012313 (2009).

\bibitem{ZavattaScience}
A. Zavatta, S. Viciani, and M. Bellini, Science {\bf 306}, 660 (2004).

\bibitem{ExpdeGauss1}
A. I. Lvovsky and S. A. Babichev, Phys. Rev. A {\bf 66}, 011801 (2002).

\bibitem{ExpdeGauss2}
J. Wenger, R. Tualle-Brouri, and P. Grangier, Phys. Rev. Lett. {\bf 92}, 153601 (2004).

\bibitem{Grangier}
A. Ourjoumtsev, A. Dantan, R. Tualle-Brouri, and P. Grangier, Phys. Rev. Lett. {\bf 98}, 030502 (2007).

\bibitem{BelliniProbing}
V. Parigi, A. Zavatta, M. Kim, and M. Bellini, Science {\bf 317}, 1890 (2007).

\bibitem{GrangierCats}
A. Ourjoumtsev, H. Jeong, R. Tualle-Brouri, and P. Grangier, Nature {\bf 448}, 784 (2007).

\bibitem{ExtremalGaussian}
M. M. Wolf, G. Giedke, and J. I. Cirac, Phys. Rev. Lett. {\bf 96}, 080502 (2006).

\bibitem{Opatrny}
T. Opatrn\'y, G. Kurizki, and D.-G. Welsch, Phys. Rev. A {\bf 61}, 032302 (2000).

\bibitem{Cochrane}
P. T. Cochrane, T. C. Ralph, and G. J. Milburn, Phys. Rev. A {\bf 65}, 062306 (2002).

\bibitem{Olivares}
S. Olivares, M. G. A. Paris, and R. Bonifacio, Phys. Rev. A {\bf 67}, 032314 (2003).

\bibitem{CVTelepNoi}
F. Dell'Anno, S. De Siena, L. Albano, and F. Illuminati, Phys. Rev. A {\bf 76}, 022301 (2007).

\bibitem{YangLi} Y. Yang and F.-L. Li, Phys. Rev. A {\bf 80}, 022315 (2009).

\bibitem{GenoniNonGaussy}
M. G. Genoni, M. G. A. Paris, and K. Banaszek, Phys. Rev. A {\bf 76}, 042327 (2007);
Phys. Rev. A {\bf 78}, 060303 (2008).

\bibitem{CVTelepNoisyNoi}
F. Dell'Anno, S. De Siena, L. Albano, and F. Illuminati, Eur. Phys. J. ST {\bf 160}, 115 (2008).

\bibitem{Glauber}
R. J. Glauber, Phys. Rev. A {\bf 131}, 2766 (1963).

\bibitem{VukicsnonidealTelep}
A. Vukics, J. Janszky, and T. Kobayashi, Phys. Rev. A {\bf 66}, 023809 (2002).

\bibitem{TelepChizhov}
A. V. Chizhov, L. Kn\"{o}ll, and D.-G. Welsch, Phys. Rev. A {\bf 65}, 022310 (2002);
A. V. Chizhov, JETP Lett. {\bf 80}, 711 (2004).

\bibitem{MarianCVTelep}
P. Marian and T. A. Marian, Phys. Rev. A {\bf 74}, 042306 (2006).

\bibitem{TelepGainBowen}
W. P. Bowen, N. Treps, B. C. Buchler, R. Schnabel, T. C. Ralph, T. Symul, and P. K. Lam,
IEEE J. Sel. Top. Quant. {\bf 9}, 1519 (2003).

\bibitem{TelepIde}
T. Ide, H. F. Hofmann, A. Furusawa, and T. Kobayashi, Phys. Rev. A {\bf 65}, 062303 (2002).

\bibitem{TelepTailored}
P. T. Cochrane and T. C. Ralph, Phys. Rev. A {\bf 67}, 022313 (2003).

\bibitem{BraunsteinKimble}
S. L. Braunstein and H. J. Kimble, Phys. Rev. Lett. {\bf 80}, 869 (1998).

\bibitem{TelepFormal1}
S. J. van Enk, Phys. Rev. A {\bf 60}, 5095 (1999).

\bibitem{TelepFormal2}
A. Vukics, J. Janszky, and T. Kobayashi, Phys. Rev. A {\bf 66}, 023809 (2000).

\bibitem{TelepFormal3}
H. F. Hofmann, T, Ide, and T. Kobayashi, Phys. Rev. A {\bf 62}, 062304 (2000).

\bibitem{FuruRep}
A. Furusawa and N. Takei, Phys. Rep. {\bf 443}, 97 (2007).

\bibitem{vanLoockTelep}
P. van Loock, Fortschr. Phys. {\bf 50}, 12 (2002).

\bibitem{LeonhardtRealHomoMeasur}
U. Leonhardt and H. Paul, Phys. Rev. A {\bf 48}, 4598 (1993).

\bibitem{WallsMilburn}
D. Walls and G. Milburn, \textit{Quantum Optics} (Berlin, Springer, 1994).

\bibitem{DecohReview}
A. Serafini, M. G. A. Paris, F. Illuminati, and S. De Siena, J. Opt. B: Quantum Semiclass. Opt. {\bf 7}, R19 (2005).

\bibitem{Benchmark}
S.L. Braunstein, C.A. Fuchs, and H.J. Kimble, J. Mod. Opt. {\bf 47}, 267 (2000);
K. Hammerer, M.M. Wolf, E.S. Polzik, and J.I. Cirac, Phys. Rev. Lett. {\bf 94}, 150503 (2005).

\bibitem{twomodeinputTelep}
S. Adhikari, A. S. Majumdar, and N. Nayak, Phys. Rev. A {\bf 77}, 012337 (2008).

\bibitem{GaussianOptimization}
G. Adesso and F. Illuminati, Phys. Rev. Lett. {\bf 95}, 150503 (2005);
A. Mari and D. Vitali, Phys. Rev. A {\bf 78}, 062340 (2008).

\end{thebibliography}
\end{document}